\documentclass[pra,twocolumn,tightenlines,nofootinbib,nolongbibliography]{revtex4-2}
\usepackage{bm,dcolumn,amsmath,graphicx,amsfonts,amssymb}

\usepackage{physics}
\usepackage[caption=false]{subfig} 
\usepackage[hidelinks]{hyperref} 
\usepackage[capitalise]{cleveref} 
\crefformat{footnote}{\textsuperscript{#2#1#3}} 

\usepackage{mhchem} 
\usepackage{soul} 
\usepackage{booktabs} 

\usepackage{hyperref}
\usepackage{xcolor}
\definecolor{cite}{rgb}{0.,0.,0.9}   
\hypersetup{colorlinks,linkcolor={cite},citecolor={cite},urlcolor={cite}}

\renewcommand{\v}[1]{\ensuremath{\boldsymbol{#1}}}		

\newcommand{\be}{\begin{equation}}	
\newcommand{\ee}{\end{equation}}	

\def\d{\ensuremath{{\rm d}}}

\newcommand{\smallspace}{\rule{0pt}{2.3ex}}

\newcommand{\un}[1]{\ensuremath{\,{\rm{#1}}}} 

\newcommand{\Fml}{\ensuremath{F^\text{ml}_\text{VP}}}
\newcommand{\Fel}{\ensuremath{F^\text{el}_\text{VP}}}

\newcommand{\A}{\ensuremath{\mathcal{A}}} 
\newcommand{\Azero}{\ensuremath{\A_0}} 

\begin{document}

\title{Vacuum polarization corrections to hyperfine structure 
 in many-electron atoms}

\author{J.\ C.\ Hasted}\email[]{j.hasted@uq.edu.au}
\author{C.\ J.\ Fairhall}
\author{O.\ R.\ Smits}
\author{B.\ M.\ Roberts}
\author{J.\ S.\ M.\ Ginges}\email[]{j.ginges@uq.edu.au}
\affiliation{School of Mathematics and Physics, The University of Queensland, Brisbane QLD 4072, Australia}
\date{\today}

\begin{abstract}\noindent
We perform a theoretical study of vacuum polarization corrections to the hyperfine structure in many-electron atoms. Calculations are performed for systems of interest for precision atomic tests of fundamental physics belonging to the alkali-metal atoms and singly-ionized alkaline earths. The vacuum polarization is considered in the Uehling approximation, and we study the many-body effects core relaxation, core polarization, and valence-core correlations in the relativistic framework. We find that for $s$ states, the relative vacuum polarization correction may be well-approximated by that for hydrogenlike ions, though for all other states account of many-body effects -- in particular, the polarization of the core -- is needed to obtain the correct sign and magnitude of the correction.

\end{abstract}

\maketitle

\section{Introduction}

The importance of quantum electrodynamics (QED) radiative corrections in many-electron heavy atoms has been firmly  established
~\cite{Sushkov2001,Johnson2001,GingesCs2002,Milstein2002,*Milstein2002a,*Milstein2003,Kuchiev2002a,*Kuchiev2002b,*KuchievQED03,
Sapirstein2003a,ShabaevPRL2005,FlambaumQED2005,Sapirstein2005,Shabaev2013,Ginges2016_lamb-shift,Ginges2016_VP,Tupitsyn2016,Fairhall2023,Kozlov2024}.
Atomic structure theory has reached the level of accuracy where QED contributions should be considered in the evaluation of a number of heavy-atom properties and effects.  
This includes in low-energy precision studies of physics beyond the standard model, most notably atomic parity violation, for which atomic calculations form a critical part~\cite{GingesRev2004,Safronova2018}. 
The testing and development of state-of-the-art atomic theory relies on the availability of measured  properties to benchmark against. Of these,  the hyperfine structure provides a unique test in the nuclear region, where the weak interaction acts. For this, 
an accurate account of QED corrections to the hyperfine structure is needed~\cite{Ginges2017}. It is also enabling in other areas of fundamental physics, including in the determination of nuclear magnetic moments for radioactive isotopes through comparison of theoretical and experimental hyperfine structure values~\cite{RobertsFr2020}, and in probing the distribution of nuclear magnetization (see, e.g.,~\cite{Skripnikov2020,Sanamyan2023}). While there is some, though limited, data available for QED corrections to the hyperfine structure in heavy atoms~\cite{SapCheng2003_mlvp_ns,SapCheng2006_mlvp_np1/2,SapCheng2008_mlvp_np3/2,Ginges2017}, the influence of many-body effects has not been considered.

Quantum electrodynamics radiative corrections are comprised of the vacuum polarization (VP) and self-energy (SE). While the SE dominates, the VP is also important, and both must be accounted for in accurate calculations. The one-loop VP is well-described by the Uehling potential~\cite{Uehling1935}, whereas the SE is more difficult to accurately model due to its non-locality. In many-electron atoms, one is faced with two possible approaches: either to perform rigorous QED calculations using a frozen atomic potential, or to use a simpler model for estimation of QED allowing for full account of many-body effects. It has been shown that for energies~\cite{Derevianko2004_relax,FlambaumQED2005,Ginges2016_lamb-shift} and electric dipole amplitudes~\cite{FlambaumQED2005,Fairhall2023} involving non-$s$ states, the latter approach is more appropriate, as the many-body effects may change the value of the QED correction by orders of magnitude, and even the sign.

There are several approaches that may be used to estimate QED effects in many-electron atoms, ions, and molecules (see, e.g., Refs.~\cite{Welton1948,Indelicato1987,Pyykko1998_VP-SE_ratio,PyykkoQED_2003,FlambaumQED2005,Thierfelder2010,Lowe2013,Dyall2013,Shabaev2013,Sunaga2022,Janke2024}). For energies, this includes approaches based on the recognition \cite{Pyykko1998_VP-SE_ratio,Labzowsky1999} that the ratio of the self-energy to Uehling correction in neutral atoms is the same as that in hydrogenlike ions of the same nucleus for $s$-states. 
To enable the inclusion of many-body effects in a straightforward manner, a local potential may be constructed to represent the non-local self-energy, and added to the many-body Hamiltonian (for example, relativistic Hartree-Fock) from the beginning, yielding self-energy corrections to the energies as well as to wave functions.    
Among the most accurate and widely-adopted approaches are the radiative potential method~\cite{FlambaumQED2005,Ginges2016_lamb-shift} and the model-operator approach~\cite{Shabaev2013}. 
The former is based on the construction and use of a local potential which reproduces the self-energy shifts for hydrogenlike ions. 
For the vacuum polarization, which is local, the situation is far simpler, and the Uehling potential may be readily included from the start.

As well as for determination of QED corrections to energies, the radiative potential method has been used to find the QED corrections to electric dipole matrix elements~\cite{FlambaumQED2005,RobertsActinides2013,Fairhall2023,RobertsE12023,TranTan2023,Kozlov2024}. Indeed, from the low-energy theorem, it is expected that the ``perturbed-orbital" corrections --  that arise from QED-corrected wave functions -- give the dominant QED contribution to matrix elements for operators that act at large distances~\cite{FlambaumQED2005}. This has been verified recently~\cite{Fairhall2023,Kozlov2024}, with agreement at the level of several percent or better. For other large-distance operators, the situation is expected to be similar. This is in contrast with short-range operators, such as the parity-violating weak interaction or the hyperfine interaction. Here, the QED corrections to the operator (vertex corrections) may be sizeable and dominate the QED contribution. Therefore, for short-range operators, both types of corrections must be carefully considered.   

In the nuclear region, where the nuclear Coulomb field is unscreened by atomic electrons, the wave functions of electrons of heavy atoms and ions and those of hydrogenlike ions with the same relativistic angular momentum quantum number $\kappa$ are proportional. 
For operators acting on the nucleus, the relative QED corrections to the matrix elements for valence states of a many-electron atom are then the same as those for hydrogenlike ions. 
For operators such as the hyperfine interaction that largely act in the unscreened part of the atom, the relation would be expected to hold approximately for $s$-states. For non-$s$ states, the Coulomb interaction between electrons plays an important role in communicating the hyperfine interaction from core states to the valence electrons. 

In this paper, we explore the influence of many-body effects on the vacuum polarization correction to the hyperfine structure in many-electron atoms with one electron above closed shells. We consider $s$, $p$, and $d$ states of the alkali-metal atoms \ce{Na} to \ce{Fr} and for the singly-ionized alkaline-earths \ce{Ba^+} and \ce{Ra^+}. Calculations are performed in the Uehling approximation, and we separately study the electric loop and magnetic loop corrections. We express our results in terms of relative corrections to allow ready comparison with the values for hydrogenlike ions. We show that, indeed, for non-$s$ states, account of many-body effects is needed to obtain reliable results.

\section{Theory}

\subsection{Hyperfine structure}
\label{ssec:hyperfine}

Electrons bound by the nucleus are described within the relativistic picture by the Dirac equation $(h_D - \varepsilon)\varphi = 0$, with $\varepsilon$ the binding energy and $h_D$ the Dirac Hamiltonian, 
\begin{equation}
  \label{eq:rhf}
    h_D = c\boldsymbol{\alpha}\cdot\mathbf{p} + (\beta-1)c^2 +  V_\text{nuc}(r) + V_\text{ee} \ .
\end{equation}
Here $\boldsymbol{\alpha}$ and $\beta$ are the Dirac matrices, $V_{\rm nuc}(r)$ is the nuclear Coulomb potential, $V_{\rm ee}$ is the potential formed by the atomic electrons, and $c=1/\alpha$ is the speed of light and $\alpha$ the fine-structure constant. We use atomic units $m_e=|e|=c\alpha = 1$ throughout unless otherwise stated. 
The four-component single-particle orbitals are of the form
\begin{equation}
\label{orbital}
\varphi_{n\kappa m}(\mathbf{r})=\frac{1}{r}
\left(
\begin{matrix}
f_{n\kappa}(r) \Omega_{\kappa m}(\theta,\phi)\\
ig_{n\kappa}(r) \Omega_{-\kappa m}(\theta,\phi)
\end{matrix}
\right) \ . 
\end{equation}
The indices $n,\, \kappa,\,  m$ are the principal, relativistic angular, and magnetic quantum numbers, respectively, and $\Omega_{\kappa m
}(\theta,\phi)$ is a two-component spherical spinor. 
The large and small radial components are denoted $f_{n\kappa}(r)$ and $g_{n\kappa}(r)$; we drop the indices in the following.

We model the nuclear charge density by a two-parameter Fermi  distribution, with thickness parameter $t$ taken to be $2.3$\un{fm} for all considered nuclei. The half-density radius $c$ is found from the root-mean-square radius $r_{\rm rms}$ compiled in Ref.~\cite{Angeli2013}, $c^2 \approx (5/3)r^2_{\rm rms} - (7/3)(\pi a)^2$, where $t=4a\ln 3$. 
For the electronic potential, we use the relativistic Hartree-Fock approximation, $V_{\rm ee}=V_{\rm HF}$~\cite{JohnsonBook2007}, which is the starting point for our many-body treatment. We will discuss this in more detail in Section~\ref{ssec:many-body}.

The magnetic hyperfine structure (hfs) arises from the interaction of the magnetic field produced by unpaired electrons with the magnetic moment of the nucleus. It is described by the interaction Hamiltonian 
\begin{equation}\label{eq:h_hfs}
    h_\text{hfs} = \alpha\frac{\boldsymbol{\mu}\cdot({\v n}\times\boldsymbol{\alpha})}{r^2}F(r) \ , 
\end{equation}
where $\v{n}=\v{r}/r$, $\v{\mu} = \mu \v{I}/I$, with $\v{I}$ the nuclear spin, $\mu$ the nuclear magnetic moment, and $F(r)$ accounting for the finite distribution of nuclear magnetization. 
The energy shift in the first order may be expressed in terms of the hyperfine constant $\A$, 
\begin{equation}
    \langle h_\text{hfs} \rangle = \left[F(F+1)-I(I+1)-J(J+1)\right]\A/2 \, ,
\end{equation}
where $J\, (F)$ is the total electronic (atomic) angular momentum. The hyperfine constant $\A$ quantifies the size of the hyperfine splitting between levels. 
For a point-nucleus magnetization distribution ($F(r)=1$), the hyperfine constant in the lowest order is given by
\be
    \Azero = \frac{\alpha}{m_p} \frac{g_I \kappa}{J(J+1)} \int_0^\infty dr f(r) g(r) / r^2,
\ee
where $g_I=\mu/(\mu_N I)$ is the nuclear g-factor, $\mu_N$ is the nuclear magneton, $m_p$ the proton mass, and $f(r)$ and $g(r)$ are found numerically by solving Eq.~(\ref{eq:rhf}) which includes the effects of the finite nuclear charge distribution.

The Bohr-Weisskopf (BW) effect \cite{Bohr1950,Bohr1951} gives a correction to the hyperfine constant  arising from a finite distribution of the nuclear magnetic moment.  
In this work we take the distribution $F(r)$ to correspond to that of a uniform magnetization distribution, $F(r)=(r/r_m)^3$ for $r\le r_m$, and $F(r)=1$ outside, where the magnetic radius $r_m=\sqrt{5/3}\,r_{\rm rms}$. Note that while there are better descriptions of the magnetization distribution -- based on the single-particle or more sophisticated nuclear models (see, e.g., \cite{Shabaev1994,Fujita1975}) -- our focus is on evaluation of the vacuum polarization correction, and in particular the {\it relative} correction, and so the details of the magnetization distribution is not so important. 
The total hyperfine constant may be given by
\begin{equation}
\A = \A_0+\A_{\rm BW}+\A_{\rm QED} \, ,
\end{equation}
where the first and second terms together correspond to the values at the considered many-body approximation, with the BW effect included, and the QED term is expressed as $F^{\rm QED}$ relative to those values,
\begin{equation}
\A_{\rm QED}=\frac{\alpha}{\pi}F^{\rm QED}(\A_0+\A_{\rm BW})\, .
\end{equation}
This has a similar form to the results of  Refs.~\cite{SapCheng2003_mlvp_ns,SapCheng2006_mlvp_np1/2,SapCheng2008_mlvp_np3/2,Ginges2017}.   
In this work, we consider the vacuum polarization component of this correction in the Uehling approximation, as described in the following section.

\begin{figure}[tb]
    \includegraphics[trim={0 0.3cm 0 0.3cm},clip,width=0.8\linewidth]{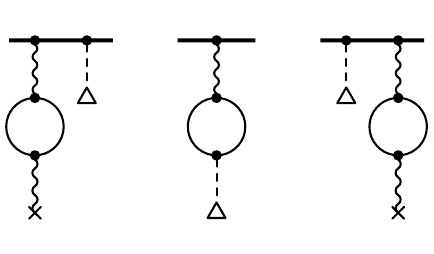}
    \caption{Feynman diagrams representing electric (left and right) and magnetic (center) VP corrections in Uehling approximation. Photon and bound (free) electron propagators are denoted by wavy and thick (thin) solid lines, respectively. Wavy line ending in cross is nuclear Coulomb interaction and dashed line ending in triangle is hyperfine interaction.}
    \label{fig:VP_feyn}
\end{figure}

\subsection{Uehling correction to hyperfine structure}
\label{ssec:uehling}

The lowest-order vacuum polarization corrections to the hyperfine structure are evaluated in the Uehling approximation, which are shown diagrammatically in Fig.~\ref{fig:VP_feyn}. There are two types of corrections: {\it electric loop}, in which the Dirac equation is solved with the Uehling potential~\cite{Uehling1935} added, and perturbed orbitals are obtained which are used to calculate the hyperfine constant; and {\it magnetic loop}, which corresponds to a vacuum polarization correction to the hyperfine operator \cite{Schneider1994}. We do not consider the higher-order-in-$Z\alpha$ Wichmann-Kroll correction~\cite{Wichmann1956_orig}, and discuss its possible effects in Section~\ref{sec:results}. 

The electric loop Uehling correction is found by evaluating the Uehling potential~\cite{Uehling1935} for a uniform nuclear charge distribution of radius {$r_n$}, 
\begin{multline} \label{eq:Vel_ueh}
    V^\text{el}_\text{Ueh}(r) = \frac{\alpha}{\pi}\frac{Z}{r}\int_1^\infty\d{t}\ \sqrt{t^2-1}\left(\frac{1}{t^2}+\frac{1}{2t^4}\right)\left(\frac{2}{\eta^3}\right) \\
    \times\begin{cases}
        \frac{r}{r_n}\eta - e^{-\eta}\left(1+\eta\right)\sinh{(2tr/\alpha)}, & r\leq r_n \\ 
        e^{-2tr/\alpha}\big[\eta\cosh{(\eta)}-\sinh{(\eta)}\big], & r > r_n
    \end{cases},
\end{multline}
where $r_n = \sqrt{5/3} r_{\rm rms}$ and $\eta=2tr_n/\alpha$  (see, e.g., Refs.~\cite{Fullerton1976,Ginges2016_VP}). 
We add the Uehling potential directly to the Dirac equation,
\begin{equation}
\label{eq:Ueh_solve}
(h_D+V_{\rm Ueh}^{\rm el})\widetilde{\varphi}=\widetilde{\varepsilon}\, \widetilde{\varphi} \ ,
\end{equation}
allowing for the inclusion of the vacuum polarization corrections to all orders \cite{indelicato2013nonperturbative}, $\widetilde{\varepsilon} = \varepsilon + \varepsilon^{(1)}+\varepsilon^{(2)}+...$ .
The electric loop correction is found by evaluating the hyperfine constant with wave functions found with and without the Uehling potential, and the difference gives the correction to the hyperfine constant, 
\begin{equation}\label{eq:F_el}
    (\alpha/\pi) \Fel{}=\frac{\mel{\widetilde{v}}{h_\text{hfs}}{\widetilde{v}}-\mel{v}{h_\text{hfs}}{v}}
{\mel{v}{h_\text{hfs}}{v}}. 
\end{equation}

On the other hand, the magnetic loop correction is found by adding a term to the hyperfine operator, 
\begin{equation}
    h_{\rm hfs} \rightarrow \widetilde{h}_{\rm hfs} = h_{\rm hfs}+h_{\rm hfs}^{\rm ml}\ .
\end{equation}
In the Uehling approximation, with the magnetization distribution taken to be uniform with extent $r_m$, it may be expressed as~\cite{volotka08a} 
\begin{multline}
\label{eq:Qmlvp}
    h^\text{ml}_\text{hfs} = \frac{\alpha^2}{\pi}\frac{\boldsymbol{\mu}\cdot(\mathbf{n}\times\boldsymbol{\alpha})}{r^2}\int_1^\infty\d{t}\ \sqrt{t^2-1}\left(\frac{1}{t^2}+\frac{1}{2t^4}\right)\frac{2}{\eta^3}\\
    \times \begin{cases}
        e^{-\eta}(1+\eta)\big[\frac{r\eta}{r_m}\cosh{\left(\frac{2tr}{\alpha}\right)}-\sinh{\left(\frac{2tr}{\alpha}\right)}\big],&r < r_m \\
        e^{-2tr/\alpha}\left(1+\frac{2tr}{\alpha}\right)\big[\eta\cosh{(\eta)}-\sinh{(\eta)}\big],&r\geq r_m
    \end{cases},
\end{multline}
with $\eta=2tr_m/\alpha$. 
The corresponding (magnetic only) correction to the hyperfine constant is found with unperturbed orbitals in evaluation of the matrix element, 
\begin{equation}
    (\alpha/\pi)\Fml{}=\langle v|h^\text{ml}_{\rm hfs}|v\rangle /\langle v|h_{\rm hfs}|v\rangle \ .
\end{equation}

We find the full Uehling correction to the hyperfine constant, including both electric and magnetic parts, by evaluating the matrix element of the magnetic-loop corrected hyperfine operator with electric Uehling-potential-perturbed orbitals, 

\begin{equation}\label{eq:F_ml}
    (\alpha/\pi)F_{\rm VP} =\frac{\mel{\widetilde{v}}{\widetilde{h}_\text{hfs}}{\widetilde{v}}-\mel{v}{h_\text{hfs}}{v}}
{\mel{v}{h_\text{hfs}}{v}}. 
\end{equation}
which absorbs the small higher-order cross terms.

\begin{figure}[!h]
    \centering
    \includegraphics[width=\linewidth]{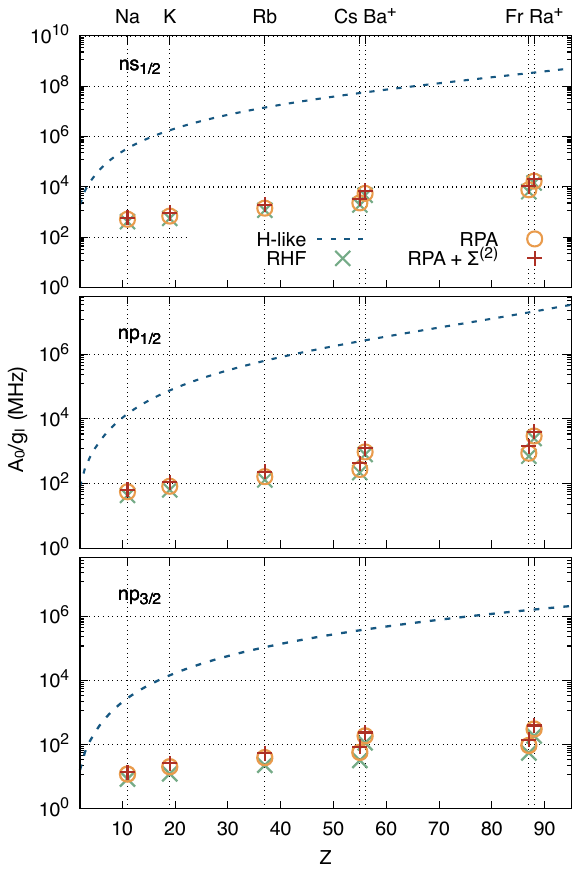}
    \caption{Zeroth-order hyperfine constants $\Azero/g_I$ in different many-body approximations. $ns_{1/2}$, $np_{1/2}$, etc. refer to ground state, lowest-excited $p_{1/2}$ state, etc. for neutral atoms and singly-ionized and H-like ions. Units: MHz.}
    \label{fig:hfs0_plot}
\end{figure}


\subsection{Many-body effects}
\label{ssec:many-body}

For the many-electron calculations, we start at the lowest order in the relativistic Hartree-Fock (RHF) approximation.
To study the significance of many-body corrections, we begin by considering the valence electron moving in the combined field of the nucleus and the ``frozen" electron core. This frozen electron potential corresponds to the RHF potential formed from the $N-1$ electrons of the atomic core, obtained from self-consistent solution of~\cref{eq:rhf}, 
\begin{equation}
    V_{\rm{ee}}=V_{\rm HF}^{N-1} \,.
\end{equation}
Equations for the Hartree-Fock potential may be found in, e.g. Ref.~\cite{JohnsonBook2007}. This is termed ``frozen" because it doesn't include the VP effects within it. To account for such ``relaxation" effects, the Uehling potential is added to the RHF equation from the beginning, and core electron orbitals and energies are found self-consistently,
\begin{equation}
    \left(h_D+V^{\rm el}_{\text{Ueh}}\right)\widetilde{\varphi}_c=\widetilde{\varepsilon}_c\widetilde{\varphi}_c \, .
\end{equation}
The valence electron orbitals and energies may then be found in the modified potential, 
\begin{equation}
    V_{\rm ee}= V_{\rm HF}^{N-1}+\delta V_{\rm relax} \ , 
\end{equation}
with core relaxation effects included; see Refs.~\cite{Derevianko2004_relax,Ginges2016_VP} for details.

Beyond RHF, there are two important many-body corrections to consider. The first of these is core polarization, in which the hyperfine field is allowed to act on the core electron orbitals, and this feeds back to the valence orbital through the electron-electron Coulomb interaction.  We take this into account using time-dependent Hartree-Fock (TDHF) method, which is equivalent to the random-phase approximation (RPA) with exchange.  The effect of this is a correction to the hyperfine operator,
\begin{equation}
\label{eq:RPA}
  {h}_{\rm hfs} \rightarrow {h}_{\rm hfs}+\delta {V}_{\rm hfs}\ ,
\end{equation}
termed the core polarization.

The second correction beyond RHF is the valence-core electron correlations. We include these in our calculations using the correlation potential method~\cite{Dzuba1984a,DzubaCPM1988pla}. The correlation potential is a non-local, energy-dependent operator which may be added to the RHF potential in Eq.~(\ref{eq:rhf}) for the valence electron. We take this potential into account in the second (lowest) order of perturbation theory, $\Sigma = \Sigma^{(2)}$. The potential is defined such that its expectation value is equal to the second-order correlation correction to the energy. By adding this potential into the RHF equation and solving for the valence orbitals and energies, the correlations are included in the solutions, 
\begin{equation}
\label{eq:Brueckner}
    (h_D + \Sigma^{(2)})\varphi^{(\rm{Br})} = \varepsilon^{(\rm{Br})}\varphi^{(\rm{Br})} \ ,
\end{equation}
termed Brueckner orbitals $\varphi^{(\rm{Br})}$ and energies $\varepsilon^{(\rm{Br})}$. These may then be used for evaluation of the hyperfine constants. 
To include core polarization and valence-core correlations together (referred to as ${\rm RPA}+\Sigma^{(2)}$ in the results section) the hyperfine constant is evaluated by taking the expectation value of the hyperfine operator \cref{eq:h_hfs} with the core polarization included through the replacement \cref{eq:RPA} and with Brueckner orbitals from \cref{eq:Brueckner}.

\section{Results and discussion}
\label{sec:results}

We illustrate in  Fig.~\ref{fig:hfs0_plot} the hyperfine constants $\Azero$ for the states $s_{1/2}$, $p_{1/2}$, and $p_{3/2}$ for H-like ions and for many-electron atoms in the RHF, RPA, and RPA+$\Sigma^{(2)}$ approximations. Due to the significantly increased overlap of the electron wave functions in the nuclear region for the H-like ions compared to many-electron atoms and ions, the corresponding hyperfine constants are orders of magnitude larger. 
It may be seen from the figure that the inclusion of core polarization and electron-core correlations typically has the effect of increasing the size of the hyperfine splitting for the many-electron atoms and ions. Indeed, core polarization allows atomic states with $J>1/2$ to gain a significant and sometimes dominating contribution that comes from participation of core $s_{1/2}$ and $p_{1/2}$ orbitals in the hyperfine interaction, while account of the valence-core correlations typically pulls in the electron orbitals, increasing the size of the effect. 
Note that in the following, our reference values will include also the BW effect, which typically enters at the level of several fractions of a percent; see, e.g., Refs.~\cite{Ginges2017,Roberts2021} for values of this effect for a number of the atoms and ions considered in this work. 

\begin{table}[!h]
    \centering
    \caption{Hyperfine constants $\Azero+\A_{\rm BW}$ and VP corrections \Fel{} and \Fml{} for \ce{^{133}Cs}. RHF includes core relaxation.
    Hyperfine constants given in GHz for H-like, and MHz otherwise.}
    \label{tab:Cs_breakdown2}
    \begin{ruledtabular}
    \begin{tabular}{lcccc}
        \ce{^{133}Cs} \smallspace & Method & $\Azero +\A_{\rm BW}$ & \Fel{} & \Fml{} \\ 
        \hline
        $ns_{1/2}$ & H-like \smallspace & 39679 & 0.862 & 0.441 \\
        & RHF \smallspace & 1421.1 & 0.930 & 0.466 \\
        & RPA \smallspace & 1712.9 & 0.928 & 0.469 \\
        & RPA+$\Sigma^{(2)}$ \smallspace & 2502.2 & 0.897 & 0.468 \\[1ex]
        $np_{1/2}$ & H-like  \smallspace & 1838.6 & 0.095 & 0.091 \\
        & RHF \smallspace & 160.62 & 0.097 & 0.098 \\ 
        & RPA \smallspace & 201.17 & 0.059 & 0.077 \\ 
        & RPA+$\Sigma^{(2)}$ \smallspace & 308.44 & 0.060 & 0.079 \\[1ex]
        $np_{3/2}$ & H-like \smallspace & 272.60 & 0.006 & 0.004  \\
        & RHF \smallspace & 23.877 & 0.007 & 0.004 \\ 
        & RPA \smallspace & 42.682 & 0.194 & 0.129 \\ 
        & RPA+$\Sigma^{(2)}$ \smallspace & 62.766 & 0.197 & 0.127 \\[1ex]
        $nd_{3/2}$ & H-like \smallspace & 49.335 & 0.000 & 0.000 \\
        & RHF \smallspace & 18.200 & -0.085 & 0.000 \\ 
        & RPA \smallspace & 15.964 & 0.925 & 0.579 \\ 
        & RPA+$\Sigma^{(2)}$ \smallspace & 47.106 & 0.681 & 0.427 \\[1ex]
        $nd_{5/2}$ & H-like \smallspace & 20.192 & 0.000 & 0.000 \\
        & RHF \smallspace & 7.4537 & -0.077 & 0.000 \\ 
        & RPA \smallspace & -24.178 & 0.754 & 0.496 \\ 
        & RPA+$\Sigma^{(2)}$ \smallspace & -44.826 & 0.866 & 0.548 \\ 
    \end{tabular}
    \end{ruledtabular}
\end{table}

\begin{figure}[b]
    \includegraphics[width=0.238\textwidth]{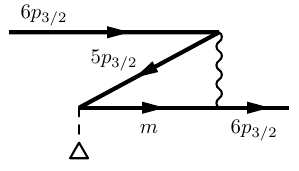}
    \includegraphics[width=0.238\textwidth]{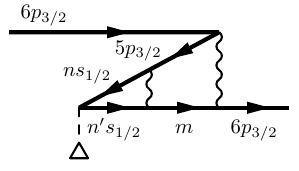}
    \caption{Example Goldstone diagrams for $5p_{3/2}$ core polarization contribution to $6p_{3/2}$ hyperfine constant, in first (left diagram) and second (right diagram) order in the Coulomb interaction. Hyperfine interaction denoted as before, and right-(left-)directed lines represent valence (core) states.}
    \label{fig:core_pol_gst}
\end{figure}

In Table~\ref{tab:Cs_breakdown2} we present our results in more detail for $s$, $p$, and $d$ states for one atom (\ce{Cs}) at different levels of many-body approximation. The H-like values are for the lowest-lying state for the specified angular momentum, e.g., $1s_{1/2}$, $2p_{1/2}$, while for the many-electron atoms it is the ground and lowest excited states, $6s_{1/2}$, $6p_{1/2}$, etc.  While core polarization and electron-core valence corrections are important for hyperfine constants for all states, their relative corrections are generally more important with increase in angular momentum. For $d_{5/2}$, the many-body corrections determine the size and sign of the hyperfine splitting. In Fig.~\ref{fig:core_pol_gst} we show, with Goldstone diagrams, in the first- and second-order of perturbation theory in the Coulomb interaction, how the core polarization effect allows significant contributions for states with valence orbitals of high angular momentum. The illustration is for the $6p_{3/2}$ state, and shows the contribution arising from the $5p_{3/2}$ core state, which in the second order opens up the hyperfine interaction involving core $s$ states. 

Results for the electric and magnetic loop vacuum polarization corrections for \ce{Cs} are given in the final columns of Table~\ref{tab:Cs_breakdown2}. 
For H-like \ce{Cs}, the relative contributions of the electric and magnetic loop VP, \Fel{} and \Fml{}, are comparable in size for each state. 
The corrections become significantly smaller with increase in orbital and total angular momentum, and for the $d$-states the contributions are smaller than the number of digits displayed in the table. 
Our H-like results are in good agreement with previously published values. For example, we obtain $\Fel{}=0.862$ and $\Fml{}=0.441$ for $1s$ for finite magnetization (Table~\ref{tab:Cs_breakdown2}), and with point magnetization we obtain $\Fel{}=0.885$ and $\Fml{}=0.454$. This may be compared to the results of Ref.~\cite{Art2001_hfs_vp} for the point-like case, which in our notations corresponds to $\Fel{}=0.883$ and $\Fml{}=0.454$, respectively. The small difference in the electric values may be due to the higher-order Uehling contribution included in our approach, Eq.~(\ref{eq:Ueh_solve}). 

For the neutral atoms, to assess the effects of electron screening on the VP results, we performed calculations in the Kohn-Sham~\cite{Kohn1965} and frozen RHF approximations, which also allowed comparison with previous values. We found excellent agreement of our Kohn-Sham results with those of Refs.~\cite{SapCheng2003_mlvp_ns,SapCheng2006_mlvp_np1/2,SapCheng2008_mlvp_np3/2,Ginges2017} for $6s_{1/2}$, $6p_{1/2}$, and $6p_{3/2}$ states. These are close in value with both our frozen RHF and H-like results, indicating, as we would expect, the factoring out of the electron screening effects for the relative corrections in these atomic approximations. Due to the similarity of these three sets of results, we have chosen only to present the H-like data, and then the influence of many-body effects may be seen relative to those values.

\begin{figure}[t]
\includegraphics[width=0.5\textwidth]{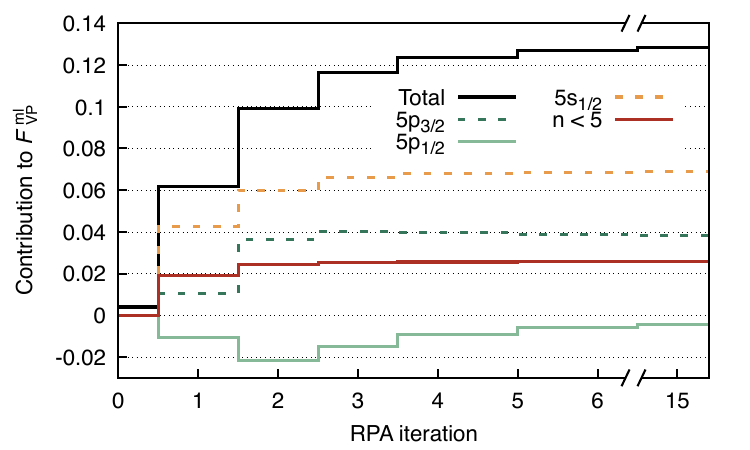}
    \caption{Contributions to \Fml{} for \ce{^{133}Cs} $6p_{3/2}$ per RPA iteration. $n<5$ refers to sum of contributions of lower orbitals. Iteration 0 refers to the RHF value.} 
\label{fig:RPA_xmag}
\end{figure}

\begin{figure}[!h]
    \centering
    \includegraphics[trim={ 0.3cm 0 0 0},clip,width=0.95\linewidth]{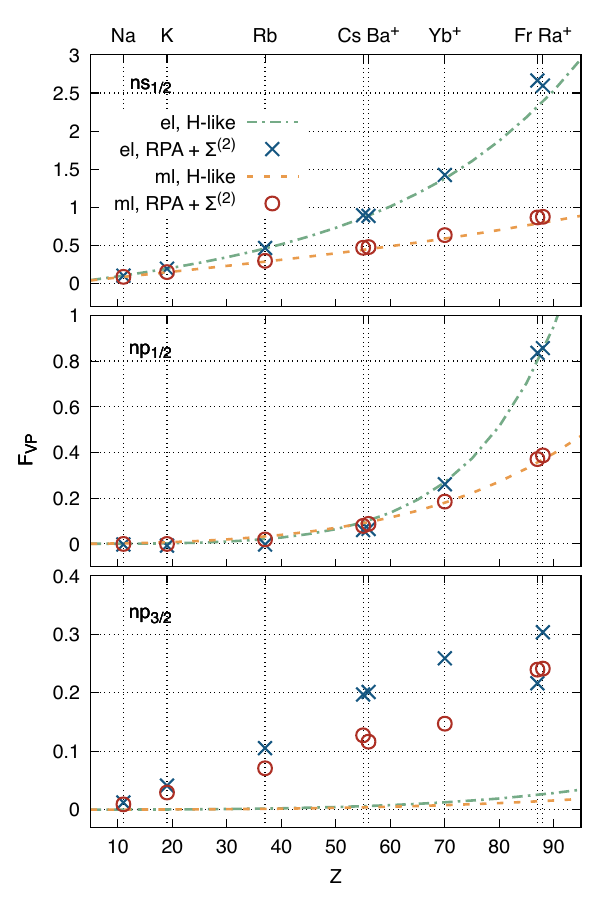}
    \caption{Relative VP corrections for many-electron alkali metal atoms and ions. Values are shown for H-like ions and many-electron atoms and ions in RPA+$\Sigma^{(2)}$ approximation.}
\label{fig:xvp_el_ml}
\end{figure}

\begin{table*}[hbt]
    \centering
    \caption{Relative electric- and magnetic-loop vacuum polarization corrections, \Fel{} and \Fml{}, and total correction $F_{\rm VP}$ for many-electron atoms in the RPA+$\Sigma^{(2)}$ approximation and their hydrogenlike counterparts.}
    \label{tab:rel_corr_breakdown}%
    \begin{ruledtabular}
    \begin{tabular}{llcccccc}
        & & \multicolumn{2}{c}{$\Fel{}$} & \multicolumn{2}{c}{$\Fml{}$} & \multicolumn{2}{c}{$F_\text{VP}$} \\ 
        \cline{3-4} \cline{5-6} \cline{7-8}  
        Atom\smallspace & State & H-like & RPA+$\Sigma^{(2)}$ & H-like & RPA+$\Sigma^{(2)}$ & H-like & RPA+$\Sigma^{(2)}$ \\ 
        \hline
        \ce{^{23}Na}\smallspace & $ns_{1/2}$ & 0.104 & 0.103 & 0.086 & 0.087 & 0.190 & 0.190 \\ 
        \smallspace & $np_{1/2}$ & 0.000 & -0.002 & 0.002 & 0.000 & 0.003 & -0.003 \\ 
        \smallspace & $np_{3/2}$ & 0.000 & 0.012 & 0.000 & 0.009 & 0.000 & 0.021 \\ 
        \smallspace & $nd_{3/2}$ & 0.000 & 0.020 & 0.000 & 0.018 & 0.000 & 0.038 \\ 
        \smallspace & $nd_{5/2}$ & 0.000 & -0.117 & 0.000 & -0.106 & 0.000 & -0.223 \\[1ex]
        \ce{^{39}K}\smallspace & $ns_{1/2}$ & 0.194 & 0.194 & 0.147 & 0.149 & 0.342 & 0.343 \\ 
        \smallspace & $np_{1/2}$ & 0.002 & -0.007 & 0.007 & 0.000 & 0.009 & -0.008 \\ 
        \smallspace & $np_{3/2}$ & 0.000 & 0.041 & 0.000 & 0.029 & 0.000 & 0.070 \\ 
        \smallspace & $nd_{3/2}$ & 0.000 & 0.258 & 0.000 & 0.213 & 0.000 & 0.472 \\ 
        \smallspace & $nd_{5/2}$ & 0.000 & 0.197 & 0.000 & 0.156 & 0.000 & 0.353 \\[1ex]
        \ce{^{87}Rb}\smallspace & $ns_{1/2}$ & 0.460 & 0.466 & 0.287 & 0.297 & 0.747 & 0.764 \\ 
        \smallspace & $np_{1/2}$ & 0.021 & -0.003 & 0.033 & 0.019 & 0.054 & 0.017 \\ 
        \smallspace & $np_{3/2}$ & 0.002 & 0.105 & 0.001 & 0.071 & 0.003 & 0.176 \\ 
        \smallspace & $nd_{3/2}$ & 0.000 & 0.543 & 0.000 & 0.391 & 0.000 & 0.935 \\ 
        \smallspace & $nd_{5/2}$ & 0.000 & 0.439 & 0.000 & 0.313 & 0.000 & 0.753 \\[1ex]
        \ce{^{133}Cs}\smallspace & $ns_{1/2}$ & 0.862 & 0.897 & 0.441 & 0.468 & 1.305 & 1.367 \\ 
        \smallspace & $np_{1/2}$ & 0.095 & 0.060 & 0.091 & 0.079 & 0.185 & 0.139 \\ 
        \smallspace & $np_{3/2}$ & 0.006 & 0.197 & 0.004 & 0.127 & 0.010 & 0.324 \\ 
        \smallspace & $nd_{3/2}$ & 0.000 & 0.681 & 0.000 & 0.427 & 0.000 & 1.110 \\ 
        \smallspace & $nd_{5/2}$ & 0.000 & 0.866 & 0.000 & 0.548 & 0.000 & 1.417 \\[1ex]     
        \ce{^{135}Ba^{+}}\smallspace & $ns_{1/2}$ & 0.890 & 0.887 & 0.451 & 0.478 & 1.343 & 1.367 \\ 
        \smallspace & $np_{1/2}$ & 0.102 & 0.065 & 0.095 & 0.088 & 0.197 & 0.153 \\ 
        \smallspace & $np_{3/2}$ & 0.006 & 0.201 & 0.004 & 0.116 & 0.010 & 0.318 \\ 
        \smallspace & $nd_{3/2}$ & 0.000 & 0.365 & 0.000 & 0.222 & 0.000 & 0.588 \\ 
        \smallspace & $nd_{5/2}$ & 0.000 & 1.707 & 0.000 & 0.961 & 0.000 & 2.673 \\[1ex]
        \ce{^{211}Fr}\smallspace & $ns_{1/2}$ & 2.323 & 2.665 & 0.788 & 0.867 & 3.118 & 3.540 \\
        \smallspace & $np_{1/2}$ & 0.797 & 0.836 & 0.356 & 0.371 & 1.155 & 1.209 \\ 
        \smallspace & $np_{3/2}$ & 0.025 & 0.216 & 0.014 & 0.239 & 0.039 & 0.456 \\ 
        \smallspace & $nd_{3/2}$ & 0.001 & 2.641 & 0.002 & 1.279 & 0.003 & 3.936 \\ 
        \smallspace & $nd_{5/2}$ & 0.000 & 1.831 & 0.000 & 0.914 & 0.000 & 2.751 \\[1ex]
        \ce{^{225}Ra^{+}}\smallspace & $ns_{1/2}$ & 2.385 & 2.597 & 0.798 & 0.875 & 3.191 & 3.481 \\ 
        \smallspace & $np_{1/2}$ & 0.844 & 0.856 & 0.368 & 0.387 & 1.214 & 1.248 \\ 
        \smallspace & $np_{3/2}$ & 0.026 & 0.303 & 0.015 & 0.241 & 0.041 & 0.545 \\ 
        \smallspace & $nd_{3/2}$ & 0.001 & 1.392 & 0.002 & 0.563 & 0.003 & 1.931 \\ 
        \smallspace & $nd_{5/2}$ & 0.000 & 2.873 & 0.000 & 1.195 & 0.000 & 4.073 \\[1ex]
    \end{tabular}
    \end{ruledtabular}
\end{table*}

The relative VP corrections for \ce{Cs} $6s_{1/2}$ at all levels of many-body approximation are very similar to those for H-like $1s_{1/2}$. 
For the $6p_{1/2}$ state, while the H-like and RHF results are similar, corrections from polarization of the core by the hyperfine field are sizeable, affecting both electric and magnetic VP corrections. 
For $J\ge 3/2$, it is seen from Table~\ref{tab:Cs_breakdown2} that core polarization determines the size and sign of the VP corrections for $F^{\rm el}_{\rm VP}$ and $F^{\rm ml}_{\rm VP}$. 
For $6p_{3/2}$, it increases the size of the (very small) RHF values by two orders of magnitude. For the $d$-states, the contributions are so sizeable that the relative VP corrections match the values for $6s_{1/2}$. It is also worth pointing out that for $d$-states, while not for $s$ or $p$ states, there is some influence from both core relaxation and from valence-core correlations.  
To illustrate in more detail the core polarization mechanism, in Fig.~\ref{fig:RPA_xmag} we show the contributions of the core states to the RPA value for $F^{\rm ml}_{\rm VP}$ for \ce{Cs} $6p_{3/2}$ as a function of iteration number. Here it is seen that at least two iterations are required to obtain a reasonable result, which is consistent with account of the second-order Goldstone diagram shown in Fig.~\ref{fig:core_pol_gst}.

Finally, in Table~\ref{tab:rel_corr_breakdown} we present our obtained relative VP results for all considered H-like ions and many-electron atoms and ions, showing the electric, magnetic, and total corrections. These values are also illustrated in Fig.~\ref{fig:xvp_el_ml}. The relative VP values for both electric and magnetic cases increase with $Z$, with the effect larger and more strongly increasing for $F_{\rm VP}^{\rm el}$ in general. 
From Fig.~\ref{fig:xvp_el_ml}, it is seen that for $s$-states the many-body results are quite close to the H-like values, though showing differences for high $Z$. For \ce{Na} and \ce{K}, the results are essentially the same, while for \ce{Fr} and \ce{Ra^+} they differ from the H-like values by 14\% and 9\%, respectively. The effect of electron screening in the many-electron systems may be seen by comparing the values for \ce{Ba^+} and \ce{Cs}, and \ce{Ra^+} and \ce{Fr} which, as expected, reduces to the values for the ions, mostly arising in the electric part. From a cursory glance at Fig.~\ref{fig:xvp_el_ml}, it may appear that the agreement between the many-body and H-like results for $p_{1/2}$ states is better than for the $s$-states. While the values agree quite well for high-$Z$ (e.g., 5\% for \ce{Fr}, and 3\% for \ce{Ra^+}), for low $Z$ this is not the case at all, and the values, while small, even have the opposite sign. 
The many-body effects are therefore important for $p_{1/2}$-states, and they become much more pronounced for $p_{3/2}$. For \ce{Fr} $7p_{3/2}$, the electric term goes against the trend seen in Fig.~\ref{fig:xvp_el_ml}, due to the influence of core polarization. Our many-body results for $p$ states are then significantly different to Kohn-Sham values presented previously~\cite{SapCheng2006_mlvp_np1/2,SapCheng2008_mlvp_np3/2}, which are close in value to the H-like data. 
For $d$ states, the many-body effects are even further pronounced, and for $d_{5/2}$ the relative corrections become larger than those for the $s$-states across all $Z$ considered here.

It may be informative to consider the influence of the Wichmann-Kroll contribution, which gives higher-order binding corrections to the vacuum polarization. 
We expect that the ratio of this contribution to Uehling for the neutral atom and H-like ion for $s$-states is roughly the same, due to their short-range nature. This assumption has been used previously to estimate the Wichmann-Kroll correction for lithiumlike \ce{Bi} -- see, e.g., Ref.~\cite{SapCheng2001_Li-like_Bi} -- with rigorous calculations in good agreement~\cite{Andreev2012}. The magnetic loop corrections are substantially larger than the electric ones, entering for Bi at about $-29\%$ and $-3\%$ the size of the Uehling corrections, respectively, and with the total Wichmann-Kroll to total Uehling ratio at about $-10\%$~\cite{Sunnergren1998,Art2001_hfs_vp}. For the atoms considered in this work, with $Z=11$ to $Z=88$, we expect this ratio to range from about $-0.2\%$ to about $-12\%$, respectively, for $s$-states based on their values for H-like ions \cite{Sunnergren1998,Art2001_hfs_vp}.   
While the size of the correction may be comparable to the many-body effects for $J=1/2$ states, we expect that its inclusion would not affect the overall behavior of what we observe for the Uehling correction. 


\section{Conclusion}

We have calculated the dominant vacuum polarization (Uehling) corrections to the hyperfine structure for a number of single-valence-electron many-electron systems of interest for precision studies of fundamental physics. We separately consider the electric and magnetic loop contributions, and study the influence of many-body effects on the relative corrections. We find that the most important many-body effect to take into account is the polarization of the atomic core by the hyperfine field. For $s$-states, one can use the relative correction for hydrogenlike ions as a good approximation for the correction in a many-electron atom or ion, though for high $Z$ there is a noticeable deviation which increases with $Z$, reaching about $10\%$ at around $Z=90$, and likely to deviate further for higher $Z$. For non-$s$-states, it is imperative to evaluate the core polarization many-body effect -- which may change the order of magnitude and even the sign of the VP correction -- in order to obtain an accurate result. 
We expect a similar trend for the more complicated and larger non-local self-energy correction.  
This work represents an important step towards considering the influence of many-body effects on the full (vacuum polarization and self-energy) QED contributions to the hyperfine structure in many-electron atoms. 

\subsection{Acknowledgments}

We thank G. Sanamyan for useful discussions. This work was supported by the Australian Government through an Australian Research Council (ARC) Discovery Project DP230101685.

\bibliography{library}

\end{document}